\pdfoutput=1
\newif\iffull\fulltrue

\documentclass[a4paper,UKenglish]{lipics-v2016}
 
\usepackage[T1]{fontenc}
\usepackage[utf8]{inputenc} 
\usepackage{microtype}

\usepackage[numbers,sort&compress]{natbib}
\usepackage{upgreek}
\usepackage{xspace,url,xcolor}
\usepackage{stmaryrd}
\usepackage{mathpartir}
\usepackage[bb=boondox]{mathalfa}
\usepackage{amsmath,amsfonts,amsthm,amssymb}
\usepackage[ruled,vlined]{algorithm2e}
\usepackage{nicefrac}
\usepackage{mathtools}
\usepackage[inline,shortlabels]{enumitem}

\usepackage{hyperref}
\hypersetup{
    unicode=true,
    pdftitle={},
    pdfauthor={},
    pdfsubject={},
    pdfnewwindow=true,
    pdfkeywords={keywords},
    colorlinks=true,
    linkcolor=black,
    citecolor=black,
    filecolor=black,
    urlcolor=black,
}

\usepackage{mathpartir}

\newcommand{\rname}[1]{[\textsc{#1}]}

\let\xinferrule\inferrule
\renewcommand{\inferrule}[3][]{\xinferrule*[right=\rname{#1}]{#2}{#3}}

\newcommand{\SYSTEM}{\mbox{\textsf{aHL}}\xspace}

\newcommand{\Sprhl}{\textsf{pRHL}\xspace}
\newcommand{\Saprhl}{\textsf{apRHL}\xspace}

\newenvironment{tightcenter}{%
  \setlength\topsep{3pt}
  \setlength\parskip{3pt}
  \begin{center}
}{%
  \end{center}
}

\DeclareMathOperator{\wt}{wt}
\DeclareMathOperator{\supp}{supp}
\DeclareMathOperator{\cmod}{mod}
\DeclareMathOperator{\FV}{FV}

\DeclareMathOperator{\opsample}{Sample}

\newcommand{\tunit}{\mathbf{unit}}
\newcommand{\tbool}{\mathbf{bool}}
\newcommand{\tint}{\mathbf{int}}
\newcommand{\treal}{\mathbf{real}}

\newcommand{\tadv}{{\mathfrak{a}}}
\newcommand{\tAdv}{{\mathfrak{A}}}

\newcommand{\farg}[1]{{#1}.\mathbf{arg}}
\newcommand{\fbody}[1]{{#1}.\mathbf{body}}
\newcommand{\fret}[1]{{#1}.\mathbf{res}}

\newcommand{\kwproc}{\mathbf{proc}}
\newcommand{\kwarg}{\mathbf{arg}}

\newcommand{\kwret}{\mathbf{res}}

\newcommand{\htx}[1]{{\lceil {#1} \rceil}}

\newcommand{\State}{\mathsf{State}}
\newcommand{\AState}{\State_{| \Adv}}

\newcommand{\varty}[1]{\tau_{#1}}
\newcommand{\osem}[1]{\overline{#1}}

\newcommand{\mnull}{{\mathbb{0}}}
\newcommand{\munit}[1]{\mathbb{1}_{#1}}
\newcommand{\mlet}[3]{\int_{#1 \leftarrow #2} {#3}}

\newcommand{\udp}[2]{[{#1} \leftarrow {#2}]}

\newcommand{\Sample}{\opsample^{\diamond}}
\newcommand{\Vars}{\ensuremath{\mathbf{Vars}}}
\newcommand{\Types}{{\ensuremath{\mathcal{T}}}}
\newcommand{\Ops}{{\ensuremath{\mathcal{O}}}}
\newcommand{\DOps}{{\ensuremath{\mathcal{O}_{\mathcal{D}}}}}

\newcommand{\Dist}{\ensuremath{\mathbf{Distr}}}

\newcommand{\Var}{\mathcal{X}}

\newcommand{\ahl}[4]{\vdash_{#4} #1 : #2 \Longrightarrow #3}

\newcommand{\denot}[1]{\llbracket #1 \rrbracket}

\newcommand{\Proc}[2]{\mathsf{proc}\; \textsc{#1}(#2):}
\newcommand{\Skip}{\mathsf{skip}}
\newcommand{\Fail}{\mathsf{abort}}
\newcommand{\Seq}[2]{{#1};\,{#2}}
\newcommand{\Ass}[2]{#1 \leftarrow #2}
\newcommand{\Rand}[2]{#1 \stackrel{\raisebox{-.25ex}[.25ex]%
  {\tiny $\mathdollar$}}{\raisebox{-.2ex}[.2ex]{$\leftarrow$}} #2}
\newcommand{\Cond}[3]{\mathsf{if}\ #1\ \mathsf{then}\ #2\ \mathsf{else}\ #3}
\newcommand{\Condt}[2]{\mathsf{if}\ #1\ \mathsf{then}\ #2}

\newcommand{\WWhile}[2]{\mathsf{while}\ #1\ \mathsf{do}\ #2}
\newcommand{\Call}[3]{#1 \leftarrow #2\mathsf{(}#3\mathsf{)}}

\newcommand{\Lap}{\mathcal{L}}

\newcommand{\Expr}{\mathcal{E}}
\newcommand{\DExpr}{\mathcal{D}}

\newcommand{\Cmd}{\mathcal{C}}
\newcommand{\qscore}{\mathsf{qscore}}
\newcommand{\Adv}{\mathcal{A}}
\newcommand{\Procs}{\mathcal{F}}
\newcommand{\FProc}[3]{\kwproc\ {#1}({\kwarg_{#1}}) \{ {#2}; \Return\ {#3}; \}}

\newcommand{\True}{\mathop{\top}}
\newcommand{\False}{\mathop{\perp}}

\usepackage[capitalise]{cleveref}

\crefname{section}{\S}{\S}
\Crefname{section}{\S}{\S}

\crefname{prop}{proposition}{propositions}
\Crefname{prop}{Proposition}{Propositions}

\crefname{lem}{lemma}{lemmas}
\Crefname{lem}{Lemma}{Lemmas}

\crefname{thm}{theorem}{theorems}
\Crefname{thm}{Theorem}{Theorems}

\crefname{definition}{definition}{definitions}
\Crefname{definition}{Definition}{Definitions}

\title{A program logic for union bounds}
\titlerunning{A program logic for union bounds} 

\author[1]{Gilles Barthe}
\author[2]{Marco Gaboardi}
\author[3]{Benjamin Grégoire}
\author[4]{\mbox{Justin Hsu}}
\author[1]{Pierre-Yves Strub}
\affil[1]{IMDEA Software Institute}
\affil[2]{University at Buffalo, SUNY}
\affil[3]{Inria Sophia Antipolis - Méditerranée}
\affil[4]{University of Pennsylvania}

\authorrunning{%
  G. Barthe and M. Gaboardi and B. Grégoire and J. Hsu and P.-Y. Strub}

\Copyright{%
  Gilles Barthe and Marco Gaboardi and Benjamin Grégoire and
  Justin Hsu and Pierre-Yves Strub}


\EventEditors{Ioannis Chatzigiannakis, Michael Mitzenmacher, Yuval
Rabani, and Davide Sangiorgi}
\EventNoEds{4}
\EventLongTitle{43rd International Colloquium on Automata, Languages,
and Programming (ICALP 2016)}
\EventShortTitle{ICALP 2016}
\EventAcronym{ICALP}
\EventYear{2016}
\EventDate{July 11--15, 2016}
\EventLocation{Rome, Italy}
\EventLogo{}
\SeriesVolume{55}
\ArticleNo{107}
\begin{document}

\maketitle

\begin{abstract}
  We propose a probabilistic Hoare logic aHL based on the union bound, a tool
  from basic probability theory. While the union bound is simple, it is an
  extremely common tool for analyzing randomized algorithms.  In formal
  verification terms, the union bound allows flexible and compositional
  reasoning over possible ways an algorithm may go wrong.  It also enables a
  clean separation between reasoning about probabilities and reasoning about
  events, which are expressed as standard first-order formulas in our logic.
  Notably, assertions in our logic are non-probabilistic, even though we can
  conclude probabilistic facts from the judgments.

  Our logic can also prove accuracy properties for interactive
  programs, where the program must produce intermediate outputs as
  soon as pieces of the input arrive, rather than accessing the entire
  input at once. This setting also enables adaptivity, where later
  inputs may depend on earlier intermediate outputs.  We show how to prove
  accuracy for several examples from the differential
  privacy literature, both interactive and non-interactive.
\end{abstract}

\section{Introduction}
Probabilistic computations arise naturally in many areas of computer science.
For instance, they are widely used in
cryptography, privacy, and security for achieving goals that lie
beyond the reach of deterministic programs.
However, the correctness of probabilistic programs can be quite subtle, often
relying on complex reasoning about probabilistic events.

Accordingly, probabilistic computations present an attractive target for formal
verification. A long line of research, spanning more than four decades, has
focused on expressive formalisms for reasoning about general probabilistic
properties both for purely probabilistic programs and for programs that combine
probabilistic and non-deterministic choice (see, e.g.,
\citep{Ramshaw79,Kozen85,Morgan96}).

More recent research investigates specialized
formalisms that work with more restricted assertions and proof
techniques, aiming to simplify formal verification.
As perhaps the purest examples of this approach,
some program logics prove probabilistic properties by
working purely with non-probabilistic assertions; we call such
systems \emph{lightweight} logics.  Examples
include \emph{probabilistic relational Hoare
logic}~\citep{BartheGZ09} for proving the reductionist security of
cryptographic constructions, and the related \emph{approximate
probabilistic relational Hoare logic}~\citep{BartheKOZ12} for
reasoning about differential privacy. These logics rely
on the powerful abstraction of \emph{probabilistic couplings}
to derive probabilistic facts from non-probabilistic
assertions~\citep{BartheEGHSS15}.

Lightweight logics are appealing because they can leverage ideas for
verifying deterministic programs, a rich and
well-studied area of formal verification.
However, existing lightweight logics apply only to
relational verification: properties about the relation between two
programs. In this paper, we propose a
non-relational, lightweight logic based on the \emph{union bound}, a
simple tool from probability theory.  For arbitrary properties
$E_1, \dots, E_n$, the union bound states that
\begin{tightcenter}
  $\Pr\left[ \cup_{i = 1}^n E_i \right] \leq \sum_{i = 1}^n \Pr[ E_i ]$ .
\end{tightcenter}

Typically, we think of the events $E_i$ as \emph{bad events}, describing
different ways that the program may fail to satisfy some target property.  Bad
events can be viewed as propositions on single program states, so they
can be represented as non-probabilistic assertions.  For
example, the formula $x > 10$ defines a bad event for $x$ a program
variable. If $x$ stores the result from a random sample, this bad event models
when the sample is bigger than $10$. The union bound states that no bad events
happen, except with probability at most the sum of the probabilities of each bad
event.

The union bound is a ubiquitous tool in pen-and-paper proofs due to its flexible
and compositional nature: to bound the probability of a collection of failures,
consider each failure in isolation. This compositional style is also a natural
fit for formal verification.  To demonstrate this, we formalize a Hoare logic \SYSTEM
based on the union bound for a probabilistic imperative language. The assertions
in our logic are non-probabilistic, but judgments carry a numeric index for
tracking the failure probability.  Concretely, the \SYSTEM judgment
\begin{tightcenter}
  $\ahl{c}{\Phi}{\Psi}{\beta}$
\end{tightcenter}
states that every execution of a program $c$ starting from an initial
state satisfying $\Phi$ yields a distribution in which $\Psi$ holds except with
probability at most $\beta$. We
define a proof system for the logic and show its soundness.
We also define a sound embedding of \SYSTEM into standard Hoare logic,
by instrumenting the program with ghost code that
tracks the index $\beta$ in a special program variable. This is a useful
reduction that also applies to other lightweight logics~\citep{BGGHKS14}.

Moreover, our logic applies both to standard algorithms and to
\emph{interactive} algorithms, a richer 
class of algorithms that is commonly studied in contexts such as
\emph{online learning} (algorithms which make predictions about the future
input) and \emph{streaming} (algorithms which operate on
datasets that are too large to fit into memory by processing the input
in linear passes). Informally, interactive algorithms receive their
input in a sequence of chunks, and must produce intermediate outputs
as soon as each chunk arrives. In some cases the input can be \emph{adaptive}:
later inputs may depend on earlier
outputs. Besides enabling new classes of algorithms, interactivity
allows more modularity. We can decompose programs
into interacting parts, analyze each part in isolation, and reuse the
components.

We demonstrate \SYSTEM on several algorithms satisfying \emph{differential
  privacy}~\citep{DMNS06}, a statistical notion of privacy which trades off
between the privacy of inputs and the accuracy of outputs. Prior work on
verifying private algorithms focuses on the privacy property for non-interactive
algorithms (see, e.g.~\citep{ReedPierce10,GHHNP13,BartheKOZ12}). We provide the
first verification of \emph{accuracy} for both non-interactive and interactive
algorithms. We note however that \SYSTEM, like the union bound, can be applied to
a wide range of probabilistic programs beyond differential privacy.

\section{A union bound logic}
Before introducing the program logic, we will begin by reviewing a largely
standard, probabilistic imperative language. We state the soundness of the logic
and describe the embedding into Hoare logic. The semantics of the language and
the proof of soundness are deferred to the appendix.

\subsection{Language}

We will work with a core imperative language with a command for random
sampling from distributions, and procedure calls. The set of commands
is defined as follows:

\begin{tightcenter}
$\begin{array}{r@{\ \ }l@{\quad}l}
\Cmd ::= & \Skip                      & \mbox{noop} \\
     \mid& \Ass{\Var}{\Expr}          & \mbox{deterministic assignment}\\
     \mid& \Rand{\Var}{\DExpr(\Expr)}     
                                      & \mbox{probabilistic assignment}\\
     \mid& \Seq{\Cmd}{\Cmd}           & \mbox{sequencing}\\
     \mid& \Cond{\Expr}{\Cmd}{\Cmd}   & \mbox{conditional}\\
     \mid& \WWhile{\Expr}{\Cmd}       & \mbox{while loop} \\
     \mid& \Call{\Var}{\Procs}{\Expr} & \mbox{procedure call} \\
     \mid& \Call{\Var}{\Adv}{\Expr}   & \mbox{external call} 

\end{array}$
\end{tightcenter}

Here, $\Var$ is a set of \emph{variables}, $\Expr$ is a set of
\emph{expressions}, and $\DExpr$ is a set of
\emph{distribution constructors}, which can be parameterized by standard
expressions.
Variables and expressions are typed, ranging over booleans, integers,
lists, etc.  The expression grammar is entirely standard, and we omit
it.

We distinguish two kinds of procedure
calls: $\Adv$ is a set of external procedure names, and $\mathcal{F}$
is a set of internal procedure names. We assume we have access to the
code of internal procedures, but not the code of external procedures. We
think of external procedures as controlled by some external
\emph{adversary}, who can select the next input in an interactive algorithm.
Accordingly, external procedures
run in an \emph{external memory} separate from the main program
memory, which is shared by all internal procedures.

For simplicity, procedures take a single argument, do not have local variables,
and are not mutually recursive.  A program consists of a sequence of procedures
definitions, each of the following form:
\begin{tightcenter}
$\FProc{f}{c}{r}$
\end{tightcenter}
Here, $f$ is a procedure name, $\kwarg_f \in \Vars$ is the formal
argument of $f$, $c$ is the function body and $r$ is its return
value. We assume that distinct procedure definitions do not bind
the same procedure name and that the program variable $\kwarg_f$ can
only appear in the body of $f$.

Before we define the program semantics, we first need to introduce a few
definitions from probability theory.

\begin{definition}
  A discrete sub-distribution over a set $A$ is defined by
  a \emph{mass function} $\mu: A \rightarrow [0, 1]$ such
  that: 
  \begin{itemize} 
    \item the support $\supp(\mu)$ of $\mu$---defined as
      $\{ x \in A \mid \mu(x) \ne 0 \}$---is countable; and

    \item the weight $\wt(\mu)$ of $\mu$---defined as $\sum_{x \in
        A} \mu(x)$---satisfies $\wt(\mu) \leq 1$.  
  \end{itemize}
  A \emph{distribution} is a sub-distribution with weight $1$.
  The probability of an event $P$ w.r.t.\ $\mu$, written $\Pr_\mu[P]$
  (or $\Pr[P]$ when $\mu$ is clear from the context), is defined as
  $\sum_{x \in A \mid P(x)} \mu(x)$.
  When $\Phi$ is an assertion (assuming that $A \equiv \State$), we
  write $\Pr_\mu[\Phi]$ for $\Pr_\mu[\lambda m .\, m \models
  \Phi]$. Likewise, when $v \in A$, we write $\Pr_\mu[v]$ for
  $\Pr_\mu[\lambda x .\, x = v]$.
\end{definition}

Commands are interpreted as a function from memories to sub-distributions
over memories, where memories are finite maps from program and external
variables to values. 
More formally, if $\State$ is the set of memories then the
interpretation of $c$, written $\denot{c}$, is a function from
$\State$ to $\Dist(\State)$, where $\Dist(\mathsf{T})$ denotes the set of
discrete sub-distributions over $\mathsf{T}$. The definition of $\denot{c}$
enforces the separation between the internal and external
states---only commands performing external procedure calls can
act on the external memory.
The interpretation of external procedure calls is
parameterized by functions---one for each external procedure---of type
$\AState \rightarrow \Dist(\AState)$, where $\AState$ is the
set of memories \emph{restricted} to the external variables.
Thus, external procedures can only access the external memory.

\subsection{Logic}
Now that we have seen the programs, let us turn to the program
logic. Our judgments are similar to standard Hoare logic with an
additional numeric index representing the probability of
failure. Concretely, the judgments are of the following form:
\begin{tightcenter}
  $\ahl{c}{\Phi}{\Psi}{\beta}$
\end{tightcenter}
where $\Phi$ and $\Psi$ are first-order formulas over the
program variables representing the pre- and post-condition, respectively. We
stress that $\Phi$ and $\Psi$ are
\emph{non-probabilistic} assertions: they do not mention the
probabilities of specific events, and will be interpreted as
properties of individual memories rather than distributions over
memories. This is reflected by the validity relation for assertions:
$m \models \Phi$ states that $\Phi$ is valid in the \emph{single} memory
$m$, rather than in a distribution over memories. Similarly, $\models \Phi$
states that $\Phi$ is valid in all (single) memories.
By separating the assertions from the probabilistic features of our
language, the assertions are simpler and easier to manipulate. The
index $\beta$ is a non-negative real number (typically, from
the unit interval $[0,1]$).

Now, we can define semantic validity for our judgments. In short, the
index $\beta$ will be an upper bound on the probability that the
postcondition $\Psi$ does not hold on the output distribution,
assuming the precondition $\Phi$ holds on the initial memory.

\begin{definition}[Validity]
  A judgment $\ahl{c}{\Phi}{\Psi}{\beta}$ is \emph{valid} if for
  every memory $m$ such that $m \models \Phi$, we have:
  \begin{tightcenter}
  $\Pr_{\denot{c}(m)}[\neg \Psi] \leq \beta$.
  \end{tightcenter}
\end{definition}

\begin{figure}[h!t]
  \begin{mathpar}
    \inferrule[Skip]
    { }
    { \ahl{\Skip}{\Phi}{\Phi}{0} }

    \inferrule[Assn]
    { }
    { \ahl{\Ass{x}{e}}{\Phi[e/x]}{\Phi}{ 0}}

    \inferrule[Rand]
    { \forall m .\, m \models \Phi \implies
        \textstyle\Pr_{\denot{\Rand{x}{d(e)}}(m)}[\neg \Psi] \le \beta }
    { \ahl{\Rand{x}{d(e)}}{\Phi}{\Psi}{\beta} }

    % \inferrule[Rand${}_0$]
    % { \forall m \models \Phi .\,
    %     \forall \xi \in \supp(\denot{\Rand{x}{d(e)}}(m)) .\,
    %       \xi \models \Psi }
    % { \ahl{\Rand{x}{d(e)}}{\Phi}{\Psi}{0} }

    \\\\
    
    \inferrule[Seq]
    { \ahl{c}{\Phi}{\Phi'}{\beta}
      \\\\
      \ahl{c'}{\Phi'}{\Phi''}{\beta'} }
    { \ahl{\Seq{c}{c'} }{\Phi}{\Phi''}{\beta + \beta'} }

    \inferrule[If]
    { \ahl{c}{\Phi \land e}{\Psi}{\beta}
      \\\\
      \ahl{c'}{\Phi \land \neg e}{\Psi}{\beta} }
    { \ahl{ \Cond{e}{c}{c'} }{\Phi}{\Psi}{\beta} }

    \inferrule[While]
    { 
      e_v : \mathbb{N}
      \\
      \models \Phi \land e_v \leq 0 \to \neg e
      \\
      \ahl{c}{\Phi}{\Phi}{\beta}
      \\
      \forall \eta > 0 .\, \ahl{c}{\Phi \land e \land e_v = \eta}{e_v < \eta}{0}
    }
    { \ahl{ \WWhile{e}{c} }{\Phi \land e_v \leq k}{\Phi \land \neg e}{k \cdot \beta} }

    \inferrule[Call]
    { \FProc{f}{c}{r}
      \\\\
      \ahl{c}{\Phi}{\Psi[r/\kwret_f]}{\beta} }
    { \ahl{ \Call{x}{f}{e}}{\Phi[e/\kwarg_f]}{\Psi[x/\kwret_f]}{\beta} }

    \inferrule[Ext]
    {~}
    { \ahl{ \Call{x}{f}{e}}{\forall v.~\Psi[v/x]}{\Psi}{0} }

    \inferrule[Weak]
    { \models \Phi' \to \Phi
      \\
      \models \Psi \to \Psi'
      \\
      \beta \leq \beta'
      \\\\
      \ahl{c}{\Phi}{\Psi}{\beta}
    }
    { \ahl{c}{\Phi'}{\Psi'}{\beta'} }

    \inferrule[Frame]
    { \mbox{$c$ does not modify variables in $\Phi$} }
    { \ahl{c}{\Phi}{\Phi}{0} }

    \inferrule[And]
    { \ahl{c}{\Phi}{\Psi}{\beta}
      \\\\
      \ahl{c}{\Phi}{\Psi'}{\beta'} }
    { \ahl{c}{\Phi}{\Psi \land \Psi'}{\beta + \beta'} }

    \inferrule[Or]
    { \ahl{c}{\Phi}{\Psi}{\beta}
      \\\\
      \ahl{c}{\Phi'}{\Psi}{\beta} }
    { \ahl{c}{\Phi \lor \Phi'}{\Psi}{\beta} }

    \inferrule[False]
    {~}
    { \ahl{c}{\Phi}{\bot}{1} }
  \end{mathpar}
  \caption{Selected proof rules.}
  \label{fig:rules}
\end{figure}

We present the main proof rules of our logic in \Cref{fig:rules}. The rule for
random sampling \rname{Rand} allows us to assume a proposition $\Psi$
about the random sample provided that $\Psi$ fails with probability at most
$\beta$. This is a semantic condition which we introduce as an axiom for
each primitive distribution.

The remaining rules are similar to the standard Hoare logic rules, with
special handling for the index. The sequence rule
\rname{Seq} states that the failure probabilities of the two commands
add together; this is simply the union bound internalized in our logic. The
conditional rule \rname{If} assumes that the indices for the two branch
judgments are equal---which can always be achieved via weakening---keeping the same
index for the conditional. Roughly, this is because only one branch of the
conditional is executed. The loop rule \rname{While} simply accumulates
the failure probability $\beta$ throughout the iterations; the side conditions
ensure that the loop terminates in at most $k$ iterations except with
probability $k \cdot \beta$. To reason about procedure calls,
standard (internal)
procedure calls use the rule \rname{Call}, which substitutes the
argument and return variables in the pre- and post-condition, respectively.
External procedure calls use the rule \rname{Ext}. We do not have
access to the implementation of the procedure; we know just the type of the
return value.

The structural rules are also similar to the typical Hoare logic
rules. The weakening rule \rname{Weak} allows strengthening
the precondition and weakening the postcondition as usual, but also
allows increasing the index---this corresponds to allowing a possibly
higher probability of failure.
The frame rule \rname{Frame} preserves assertions that do not mention
variables modified by the command. The conjunction rule \rname{And} is
another instance of the union bound, allowing us to combine two
postconditions while adding up the failure probabilities. The case
rule \rname{Or} is the dual of \rname{And} and takes the
maximum failure probability among two post-conditions when taking their
disjunction. Finally, the rule \rname{False} allows us to conclude
false with failure probability $1$: With probability at most $0$,
false holds in the final memory.

We can show that our proof system is sound with respect to the semantics; the
proof is deferred to the appendix.
\begin{theorem}[Soundness]\label{th:soundness}
  All derivable judgments $\ahl{c}{\Phi}{\Psi}{\beta}$ are valid.
\end{theorem}

In addition, we can define a sound embedding into Hoare logic in the
style of~\citet{BGGHKS14}. Assuming a fresh program variable
$x_{\beta}$ of type $\mathbb{R}$, we can transform a command
$c$ such that $\ahl{c}{\Phi}{\Psi}{\beta}$ to a new command $\htx{c}$ and
a proof of the standard Hoare logic judgment
\begin{center}
  $\vdash \htx{c} : \Phi \land x_{\beta} = 0 \implies
     \Psi \land x_{\beta} \le \beta$ .
\end{center}
The command $\htx{c}$ is obtained from $c$ by replacing all probabilistic
sampling $\Rand{x}{d(e)}$ with a call to an abstract, non-probabilistic
procedure call $\Call{x}{\Sample}{d(e)}$, whose specification models the
postcondition of \rname{Rand}:
\begin{center}
  $\xinferrule
     {\forall m .\, m \models \Phi \implies
       \Pr_{\denot{\Rand{x}{d(e)}}(m)}[\neg \Psi] \le \iota}
     {\vdash \Call{x}{\Sample}{d(e)} :
       \Phi \land x_{\beta} \leq \nu \implies \Psi \land x_{\beta} \le \nu +
       \iota}$ .
\end{center}

\section{Accuracy for differentially private programs}

Now that we have presented our logic \SYSTEM, we will follow by verifying
several examples. Though our system applies to programs from many domains, we will
focus on programs satisfying \emph{differential privacy}, a
statistical notion of privacy proposed by \citet{DMNS06}. At a very
high level, these programs take private data as input and add random
noise to protect privacy. (Interested readers should consult a
textbook~\citep{DR14} for a more detailed presentation.) In
contrast to existing formal verification work, which verifies the
privacy property, we will verify \emph{accuracy}. This is just as
important as privacy: the constant function is perfectly private but not very
useful.

All of our example programs take samples from the Laplace distribution.
\begin{definition}
The \emph{(discrete) Laplace} distribution $\Lap_\epsilon(e)$ is parameterized
by a scale parameter $\epsilon > 0$ and a mean $e$. The
distribution ranges over the real numbers $\{ \nu = k + e \}$ for $k$ an integer,
releasing $\nu$ with probability proportional to:

\begin{tightcenter}
  $\Pr_{\Lap_\epsilon(e)}[ \nu ] \propto \exp{(-\epsilon \cdot |\nu - e|)}$.
\end{tightcenter}
\end{definition}

This distribution satisfies a basic accuracy property.

\begin{lemma}
  Let $\beta \in (0, 1)$, and let $\nu$ be a sample from the distribution
  $\Lap_\epsilon(e)$. Then,
  \begin{tightcenter}
    $\Pr_{\Lap_\epsilon(e)} \left[
         |\nu - e| > \frac{1}{\epsilon} \log \frac{1}{\beta}
     \right] < \beta$ .
  \end{tightcenter}
\end{lemma}

Thus, the following sampling rule is sound for our system for every 
$\beta \in (0, 1)$:
\begin{tightcenter}
  $\inferrule[LapAcc]
  {~}
  { \ahl{\Rand{x}{\Lap_\epsilon(e)}}
    {\top}
    { |x - e| \leq \frac{1}{\epsilon} \log \frac{1}{\beta} }
    {\beta} }$
\end{tightcenter}
% For $\beta=0$, we take the trivial specification 
% $\ahl{\Rand{x}{\Lap_\epsilon(e)}}{\top}{\top}{0}$.

Before presenting the examples, we will set some common notations and terminology.
First, we consider a set \textsf{db} of databases,\footnote{%
  The general setting of differential privacy is that the database contains
  private information that must be protected.  However, this fact will not be
  important for proving accuracy.}
a set \textsf{query} of queries, and primitive functions
$$\begin{array}{rcl}
\textsf{evalQ} & : & \textsf{query}\rightarrow \textsf{db}\rightarrow
\mathbb{R} \\
\textsf{invQ} & : & \textsf{query}\rightarrow \textsf{query} \\
\textsf{negQ} & : & \textsf{query}\rightarrow \textsf{query} \\
\textsf{size} & : & \textsf{db}\rightarrow \mathbb{N} \\
\textsf{error} & : &\textsf{query}\rightarrow \textsf{db} \rightarrow
\textsf{query}
\end{array}$$
satisfying 
$$\begin{array}{rcl}
\textsf{evalQ}(\textsf{invQ} (q),d) & = & -\textsf{evalQ}(q,d) \\
\textsf{evalQ}(\textsf{negQ} (q), d) & = & \textsf{size}(d) -\textsf{evalQ}(q,d) \\
\textsf{evalQ}(\textsf{error}(q, d_1), d_2) & = &
\textsf{evalQ}(q,d_1)-\textsf{evalQ}(q,d_2)
\end{array}$$
Concretely, one can identify \textsf{query} with the functions
$\textsf{db}\rightarrow\mathbb{R}$ and obtain an easy realization of the
above functions and axioms.

In some situations, we may need additional structure on the queries to prove the
accuracy guarantees. In particular, a query $q$ is \emph{linear} if
\begin{itemize}
  \item for every two databases $d, d'$, we have $q(d + d') = q(d) + q(d')$ for
    a commutative and associative operator $+$ on databases; and
  \item for the database $d_0$ that is the identity of $+$, we have $q(d_0) = 0$.
\end{itemize}
Concretely, we can identify $\textsf{db}$ with the set of multisets, $+$
with multiset union, and $d_0$ with the empty multiset.

\subsection{Report-noisy-max}
Our first example is the \emph{Report-noisy-max} algorithm (see, e.g.,
\citet{DR14}). Report-noisy-max is a variant of the  \emph{exponential 
mechanism}~\citep{MT07}, which provides the standard way to achieve
differential privacy for computations whose outputs lie in a finite (perhaps
non-numeric)
set $\mathcal{R}$. Both algorithms perform the same computations,
except that the exponential mechanism adds \emph{one-sided} Laplace
noise whereas Report-noisy-max adds regular Laplace noise. Thus,
accuracy for both algorithms is verified in essentially the same
way. We focus on Report-noisy-max to avoid defining
one-sided Laplace.

Report-noisy-max finds an element of a finite set $\mathcal{R}$ that
approximately maximizes some \emph{quality score} function $\qscore$,
which takes as input an element $r \in \mathcal{R}$ and a database $d$.
Operationally, Report-noisy-max computes the quality score for each
element of $\mathcal{R}$, adds Laplace noise, and returns the
element with the highest (noisy) value. We can implement this
algorithm with the following code, using syntactic sugar for arrays:
\[
  \begin{array}{l}
    \Proc{RNM}{\mathcal{R}, d} \\
    \quad \Ass{\mathit{flag}}{1}; \Ass{\mathit{best}}{0}; \\
    \quad \WWhile{\mathcal{R}\neq \emptyset}{} \\
    \quad \quad \Ass{r}{\mathsf{pick}(\mathcal{R})};
    \Rand{\mathit{noisy}[r]}{\Lap_{\epsilon/2}(\qscore(r,d))};  \\
    \quad \quad \mathsf{if}~ (\mathit{noisy}[r] > \mathit{best} \vee \mathit{flag}=1)
    ~\mathsf{then} \\
    \quad \quad \quad \Ass{\mathit{flag}}{0}; \Ass{r^*}{r}; \Ass{\mathit{best}}{\mathit{noisy}[r]}; \\
    \quad \quad \Ass{\mathcal{R}}{\mathcal{R}\setminus \{ r \};} \\
    \mathsf{return}~r^*;
  \end{array}
\]
The scale $\epsilon/2$ of the Laplace distribution ensures an appropriate level
of differential privacy under certain assumptions; we will not discuss privacy
in the remainder.
\begin{theorem} \label{thm:rnm-acc}
  Let $\beta \in (0, 1)$, and let $\mathit{res} \in \mathcal{R}$ be the output
  of Report-noisy-max on input $d$ and quality score $\qscore$. Then,
  we have the following judgment:
  \begin{tightcenter}
    $\ahl{\textsc{RNM}}{\top}
      {\forall r \in \mathcal{R}.\; \qscore(\mathit{res}, d) >
        \qscore(r, d) -
          \frac{4}{\epsilon} \log \frac{ |\mathcal{R}| }{ \beta } }
      {\beta}$.
  \end{tightcenter}
  where $|\mathcal{R}|$ denotes the size of $\mathcal{R}$. This corresponds to
  the existing accuracy guarantee for Report-noisy-max
  (see, e.g., \citet{DR14}).
\end{theorem}
Roughly, this theorem states that while the result $\mathit{res}$ may not be
the element with the absolute highest quality score, its quality score
is not far below the quality score of any other element. For a brief
sanity check, note that the guarantee weakens as we increase the
range $\mathcal{R}$, or decrease the failure probability $\beta$.

The proof of accuracy is based on an instantiation of the
rule \rname{LapAcc} with $e$ set to $\qscore(r,d)$,
$\beta$ set to $\beta/|\mathcal{R}|$, and $\epsilon$ set to $\epsilon/2$. First,
we can show
\begin{tightcenter}
  $\ahl{c}
  {\top} {|\mathit{noisy}[r] - \qscore(r, d)| < \frac{2}{\epsilon} \log
      \frac{ | \mathcal{R}| }{ \beta } }
  {\beta/| \mathcal{R}|}$
\end{tightcenter}
where $c$ is the loop body. Since the loop runs for $| \mathcal{R}|$ 
iterations, we also have
\begin{tightcenter}
  $\ahl{\textsc{RNM}}{\top}
    {\forall r \in \mathcal{R}.\; |\mathit{noisy}[r] - \qscore(r, d)|
      < \frac{2}{\epsilon} \log \frac{ | \mathcal{R}| }{ \beta } }
    {\beta}$.
\end{tightcenter}
In order to prove this judgment, the loop invariant
quantifies over all previously seen $r\in\mathcal{R}$.  Combined with a
straightforward invariant showing that $r^*$ stores the index of the current
maximum (noisy) score, the above judgment suffices to prove the accuracy
guarantee for Report-noisy-max (\Cref{thm:rnm-acc}).

\subsection{Sparse Vector algorithm}

Our second example is the \emph{Sparse Vector algorithm}, which indicates
which numeric queries take value (approximately) above
some threshold value (see, e.g., \citet{DR14}). Simpler approaches can
accomplish this task by releasing the noisy answer to all queries and then
comparing with the threshold, but
the resulting error then grows linearly with the total number of
queries. Sparse Vector does not release the noisy answers, but the resulting
error grows only \emph{logarithmically} with the total number of queries---a
substantial improvement. The differential privacy property of Sparse Vector was
recently formally verified~\citep{BGGHS16}; here, we consider the accuracy
property.

In the non-interactive setting, the algorithm takes as input a list of
queries $q_1, q_2,\dots$, a database $d$, and a numeric threshold $t
\in \mathbb{R}$.\footnote{%
  In some presentations, the algorithm is also parameterized by the maximum
  number $k$ of queries to answer. This feature is important for privacy but not
  accuracy, so we omit it. It is not difficult to extend the accuracy proof for
  answering at most $k$ queries.}
First, we add Laplace noise to the threshold $t$ to
calculate the noisy threshold $T$. Then, we evaluate each query $q_i$ on
$d$,
add Laplace noise, and check if the noisy value exceeds $T$. If so, we output
$\top$; if not, we output $\bot$.

Sparse Vector also works in the
\emph{interactive} setting. Here, the algorithm is fed one query at
a time, and must process this query (producing $\bot$ or $\top$)
before seeing the next query. The input may be
adaptive---future queries may depend on the answers to earlier
queries.

We focus on the interactive version; the non-interactive version can be handled
similar to Report-noisy-max. We break the code into two pieces. The
first piece initializes variables and computes the noisy
threshold, while the second piece accepts a single new query and returns the
answer.
\begin{mathpar}
  \begin{array}{l}
    \Proc{SV.Init}{T_{in}, \epsilon_{in}} \\
    \quad \Ass{\epsilon}{\epsilon_{in}}; \\
    \quad \Rand{T}{\Lap_{\epsilon/2}(T_{in})}; \\
  \end{array}

  \begin{array}{l}
    \Proc{SV.Step}{q} \\
    \quad \Rand{a}{\Lap_{\epsilon/4}(\mathsf{evalQ}(q, d))};  \\
    \quad \mathsf{if}~(a < T)~\mathsf{then}~\{ \Ass{z}{\bot}; \}
    ~\mathsf{else}~\{ \Ass{z}{\top}; \} \\
    \mathsf{return}~z;
  \end{array}
\end{mathpar}
% Again, the precise privacy parameters involving $\epsilon$ will not be
% important for us.
The main procedure performs initialization, and then enters into an interactive
loop between the external procedure $\mathcal{A}$---which supplies the
queries---and the Sparse Vector procedure \textsc{SV.Step}:
\[
  \begin{array}{l}
    \Proc{SV.main}{Q, T, \epsilon} \\
    \quad \textsc{SV.Init}(T, \epsilon); \\
    \quad \Ass{u}{0}; \Ass{ans[u]}{\bot}; \\
    \quad \WWhile{(u < Q)}{} \\
    \quad\quad \Ass{u}{u + 1}; \\
    \quad\quad \Ass{q[u]}{\mathcal{A}(ans[u - 1])}; \\
    \quad\quad \Ass{ans[u]}{\textsc{SV.Step}(q[u])}; \\
    \mathsf{return}~ans;
  \end{array}
\]
\pagebreak
Sparse Vector satisfies the following accuracy guarantee.
\begin{theorem} \label{thm:sv-acc}
  Let $\beta \in (0, 1)$. We have
  \begin{center}
    $\ahl{\textsc{SV.main}(Q, T)}{\top}
      { \forall j \in \{ 1, \dots, Q \}.\; \Phi(q[j], d) }{\beta}$, where
  % \end{center}
  % \begin{center}
    \begin{align*}
      \Phi(q, d) \triangleq &
      \left( \mathit{res} = \top \to \mathsf{evalQ}(q, d) > t
        - \frac{6}{\epsilon} \log \frac{Q + 1}{\beta} \right) \\
      &\land
      \left( \mathit{res} = \bot \to \mathsf{evalQ}(q, d) < t
        + \frac{6}{\epsilon} \log \frac{Q + 1}{\beta} \right) .
    \end{align*}
  \end{center}
  This judgment corresponds to the accuracy guarantee for Sparse
  Vector from (see, e.g., \citet{DR14}). Note that the error term depends
  logarithmically on the total number of queries $Q$, a key feature of Sparse
  Vector.
\end{theorem}

To prove this theorem, we first specify the procedures \textsc{SV.Init} and
\textsc{SV.Step}. For initialization, we have

\begin{tightcenter}
  $\ahl{\textsc{SV.Init}(T, \epsilon)}{\top}{\Phi_t}{\beta/(Q + 1)}$
  $\quad$ where $\quad$
  $\Phi_t \triangleq |t - T| < \frac{2}{\epsilon} \log \frac{Q +
      1}{\beta}
    \land \epsilon = \epsilon_{in}$ .
\end{tightcenter}
For the interactive step, we have
\begin{tightcenter}
  $\ahl{\textsc{SV.Step}(q)}{\Phi_t}{\Phi_t \land \Phi(q, d)}{\beta/(Q + 1)}$ .
\end{tightcenter}
Combining these two judgments, we can prove accuracy for
\textsc{SV.main} (\Cref{thm:sv-acc}).

\subsection{Online Multiplicative Weights}

Our final example demonstrates how we can use the union bound to analyze a
complex
combination of several interactive algorithms, yielding sophisticated accuracy
proofs. We will verify the \emph{Online Multiplicative Weights} (OMW) algorithm
first proposed by \citet{HR10} and later refined by \citet{GRU12}.
Like Sparse Vector, this interactive algorithm can handle
adaptive queries while guaranteeing error logarithmic in the number of queries.
Unlike Sparse Vector, OMW produces approximate answers
to the queries instead of just a bit representing above or below threshold.

At a high level, OMW
iteratively constructs a \emph{synthetic} version of the true database. The user
can present
various linear queries to the algorithm, which applies the Sparse Vector algorithm to
check whether the error of the synthetic database on this query is smaller than
some threshold.
If so, the algorithm simply returns the approximate answer.
Otherwise, it updates the
synthetic database using the \emph{multiplicative weights} update rule to
better model the true database, and answers the query by adding Laplace noise to
the true answer. An inductive argument shows that after enough
updates, the synthetic database must be similar to the true database
on \emph{all} queries. At this point, we can answer all subsequent queries using
the synthetic database alone.

In code, the following procedure implements the Online Multiplicative Weights
algorithm.
\[
  \begin{array}{ll} 
    \Proc{MW-SV.main}{d, \alpha, \epsilon, Q, X, n} & \\
    \quad \Ass{\eta}{\alpha/2n};
    \Ass{T}{2\alpha}; 
    \Ass{c}{4 n^2 \ln(X)/\alpha^2}; & \text{set parameters}\\
    \quad \Ass{u}{0};
    \Ass{k}{0};
    \Ass{\mathit{ans}[k]}{\bot}; & \text{initialize variables} \\
    \quad \Ass{\mathit{mwdb}}{\textsc{MW.Init}(\eta, X, n)};
    \textsc{SV.Init}(T, \epsilon/4c); & \text{initialize MW and SV} \\
    \quad \WWhile{(k < Q)}{} & \text{main loop} \\
    \quad\quad \Ass{k}{k + 1}; & \text{increment count of queries} \\
    \quad\quad \Ass{q[k]}{\Adv(\mathit{ans}[k - 1], \mathit{mwdb})}; & \text{get next query} \\
    \quad\quad \Ass{\mathit{approx}}{\mathsf{evalQ}(q[k], \mathit{mwdb})}; &
    \text{calculate approx answer}\\
    \quad\quad \Ass{\mathit{exact}}{\mathsf{evalQ}(q[k], d)}; & 
    \text{calculate exact answer}\\
    \quad\quad \mathsf{if}~(k \geq c)~\mathsf{then}~
    \Ass{\mathit{ans}[k]}{\mathit{approx}};
    & \text{enough updates, use approx answer} \\
    \quad\quad \mathsf{else} \\
    \quad\quad\quad \Ass{\mathit{err}_>}{\textsf{error}(q[k], \mathit{mwdb})};
    \Ass{\mathit{at}}{\textsc{SV.Step}(\mathit{err}_>)}; &
    \text{check if approx answer is high} \\
    \quad\quad\quad \Ass{\mathit{err}_<}{\textsf{invQ}(\textsf{error}(q[k],
      \mathit{mwdb}))};
    \Ass{\mathit{bt}}{\textsc{SV.Step}(\mathit{err}_<)}; &
    \text{check if approx answer is low} \\
    \quad\quad\quad \mathsf{if}~(at \neq \bot \lor bt \neq \bot)~\mathsf{then} &
    \text{large error} \\
    \quad\quad\quad\quad
    \Ass{u}{u + 1}; & \text{increment count of updates} \\
    \quad\quad\quad\quad
    \textsf{if}~{at \neq \bot}~\textsf{then}~\Ass{\mathit{up}}{q[k]}; & \text{approx answer too high} \\
    \quad\quad\quad\quad \textsf{else}~\Ass{\mathit{up}}{\textsf{negQ}(q[k])}; &
    \text{approx answer too low} \\
    \quad\quad\quad\quad
    \Ass{\mathit{mwdb}}{\textsc{MW.Step}(\mathit{mwdb}, \mathit{up})}; &
    \text{update synthetic db} \\
    \quad\quad\quad\quad
    \Rand{\mathit{ans}[k]}{\Lap_{\epsilon/2c} (\mathit{exact})}; & \text{estimate true answer} \\
    \quad\quad\quad \mathsf{else}~ &
    \text{small error, do not update} \\
    \quad\quad\quad\quad \Ass{ans[k]}{\mathit{approx}};
      & \text{answer using approx answer} \\
    \mathsf{return}~\mathit{ans}; &
  \end{array}
\]
% Here, we assume that each query $q[u]$ is given as a list of values in $\{ 0, 1
% \}$, and is evaluated on database $db$ by taking the dot product $\langle q[u],
% db \rangle$. The database will be a list of natural numbers summing up to $n$.
% The program uses both the query $q[u]$ and its negated version $1 - q[u]$
% (flipping each entry of $q[u]$).

% We also use the conditional expression $e~?~e_t:e_f$, which is
% interpreted as $e_t$ if $e$ is true, and $e_f$ if $e$ is false.

Online multiplicative weights satisfies the following accuracy guarantee.

\begin{theorem} \label{thm:mw-acc}
  Let $\beta \in (0, 1)$. Then,

  \begin{tightcenter}
    $\begin{array}{l}
      \ahl{\textsc{MW-SV.main}(d, \alpha, \epsilon, Q, X, n)}
      {\alpha \geq \max(\alpha_{sv}, \alpha_{lap})}{}
      {\beta} \\[.35em]
      \quad \quad
      \forall j.\; j \in \{ 1, \dots, Q \} \to
      | \mathit{res}[j] - \mathsf{evalQ}(q[j], d) | \leq \alpha ,
    \end{array}$
  \end{tightcenter}
  %
  % where
  %
  \begin{tightcenter}
    $\begin{array}{l}
      \text{where}
      \quad
      \gamma \triangleq 4n^2 \ln(X)/\alpha^2 ,
      \quad
      \alpha_{sv} \triangleq \frac{24\gamma}{\epsilon} \log \frac{2(Q +
        1)}{\beta} ,
      \quad
      \text{and}
      \quad
      \alpha_{lap} \triangleq \frac{4\gamma}{\epsilon} \log \frac{2\gamma}{\beta}.
    \end{array}$
  \end{tightcenter}
  In words, the answers to all the supplied queries are within $\alpha$ of
  the true answer if $\alpha$ is sufficiently large.  The above judgment reflects
  the accuracy guarantee first proved by \citet{HR10} and later generalized by
  \citet{GRU12}.
\end{theorem}

The main routine depends on the \emph{multiplicative weights}
subroutine (MW), which maintains and updates the synthetic
database. Roughly, MW takes as input the current synthetic database and a query
where the
synthetic database gives an answer that is far from the true answer.
Then, MW improves the synthetic database to better model the true database.
Our implementation of MW consists of two subroutines: \textsc{MW.init}
initializes the synthetic database, and \textsc{MW.step} updates the current
database with a query that has high error.
The code for these subroutines is somewhat technical, and we will not
present it here.

Instead, we will present their specifications, which are given in
terms of an expression $\Psi(x, d)$ where $x$ is the current
synthetic database and $d$ is the true database. We omit the
definition of $\Psi$ and focus on its three key properties:
\begin{itemize}
  \item $\Psi(x, d) \geq 0$;
  \item $\Psi(x, d)$ is initially bounded for the initial synthetic database;
    and
  \item $\Psi(x, d)$ decreases each time we update the synthetic database.
\end{itemize}
Functions satisfying these properties are often called \emph{potential
  functions}.

The first property follows from the definition of $\Psi$, while the second and third
properties are reflected by the specifications of the MW procedures.
Concretely, we can bound the initial value of $\Psi$ with the following
specification for \textsc{MW.init}:
\begin{tightcenter}
  $\ahl{\textsc{MW.init}(\eta, X, n)}{\top}{ \Psi(\mathit{res}, d) \leq 
\ln X }{0}$
\end{tightcenter}
We can also show that $\Psi$ decreases with the following specification for
\textsc{MW.step}:
\begin{tightcenter}
  $\ahl{\textsc{MW.step}(x, q)}{\top}{ \Psi(x, d) - \Psi(\mathit{res},
    d) \geq
    \eta( \mathsf{evalQ}(q, x) - \mathsf{evalQ}(q, d) )/n - \eta^2 }{0}$
\end{tightcenter}
We make two remarks. First, these specifications crucially rely on the fact that
$q$ is a linear query.  Second, both procedures are deterministic. For such
procedures, the fragment of \SYSTEM with index $\beta = 0$ corresponds precisely
to standard Hoare logic.

Now, let us briefly consider the key points in proving the main specification
(\Cref{thm:mw-acc}). First, the key part of the invariant for the main loop is
$\Psi(\mathit{mwdb}, d) \leq \log X - u \cdot \alpha^2/4n^2$. Roughly, $\Psi$
is initially at
most $\log X$ by the specification for \textsc{MW.init}, and every time we call
\textsc{MW.step} we decrease $\Psi$ by at least $\alpha^2/4n^2$ if the update
query $up$ has error at least $\alpha$. Since $\Psi$ is always non-negative, we can find
at most $c$ queries with high error---after $c$ updates,
the synthetic database $\mathit{mwdb}$ must give accurate answers on
all queries.

Prior to making $c$ updates, there are two cases for each query. If at least one
of the Sparse Vector
calls returns above threshold, we set the update query $\mathit{up}$ to be $q[u]$ if the
approximate answer is too high, otherwise we set $\mathit{up}$ to be the negated query
$\mathsf{neqQ}(q[u])$ if the approximate answer is too low. With this choice of
update query, we can show that
\begin{tightcenter}
  $\mathsf{evalQ}(\mathit{up}, \mathit{mwdb}) - \mathsf{evalQ}(\mathit{up},
  d) \geq \alpha$
\end{tightcenter}
so $\Psi$ decreases by at least $\alpha^2/4n^2$. Then, we
answer the original query $q[u]$ by adding Laplace noise, so our
answer is also within $\alpha$ of the true answer. Otherwise, if both
Sparse Vector calls return below threshold, then the
query $q[u]$ is answered well by our approximation $\mathit{mwdb}$ and there
is no need to update $\mathit{mwdb}$ or access the real database $d$.

The above reasoning assumes that Sparse Vector and the Laplace
mechanisms are sufficiently accurate. To guarantee the former, notice
that the Sparse Vector subroutine will process at most $2Q$ queries, so we
assume that $\alpha$ is larger than the error $\alpha_{sv}$ guaranteed
by \Cref{thm:sv-acc} for $2Q$ queries and failure probability
$\beta/2$. To guarantee the latter, notice that we sample Laplace
noise at most $c$ times---once for each update step---so we assume
that $\alpha$ is larger than the error $\alpha_{lap}$ guaranteed
by \rname{LapAcc} for failure probability $\beta/2c$; by a
union bound, all Laplace noises are accurate except
with probability $\beta/2$. Taking $\alpha \geq
\max(\alpha_{sv}, \alpha_{lap})$, both accuracy guarantees hold except with
probability at most $\beta$, and we have the desired proof of accuracy for OMW
(\Cref{thm:mw-acc}).

\section{Related work}

The semantics of probabilistic programming languages has been studied
extensively since the late 70s.
Kozen's seminal paper~\citep{Kozen79} studies two semantics
for a core probabilistic imperative language. Other important work
investigates using monads to structure the semantics of probabilistic
languages; e.g.~\citet{JonesP89}.
More recent works study the semantics of probabilistic
programs for applications like statistical
computations~\citep{Bhat:2012}, probabilistic inference for machine
learning~\citep{BGGMG13}, probabilistic modeling for software defined
networks~\citep{FKMRS16}, and more.

Likewise, deductive techniques for verifying probabilistic programs
have a long history. \citet{Ramshaw79} proposes
a program logic with basic assertions of the form $\Pr[E]=p$.  
\citet{SharirPH84, HartSP82} propose a method using intermediate assertions and
invariants for proving general properties of probabilistic programs. 
\citet{Kozen85} introduces \textsf{PPDL}, a logic that can reason
about expected values of general measurable functions.
\citet{Morgan96} (see \citet{McIverM05} for an extended account) 
propose a verification method based on computing \emph{greatest
pre-expectations}, a probabilistic analogue of Dijkstra's weakest
pre-conditions. \citet{HurdMM05} formalize their approach using the
\textsf{HOL} theorem prover. Other approaches based on interactive theorem
provers include the work of \citet{audebaud2009proofs}, who axiomatize
(discrete) probability theory and verify some examples of randomized
algorithms using the \textsf{Coq} proof assistant. \citet{GretzKM13} extend the
work of~\citet{Morgan96} with a formal treatment of conditioning. More
recently, \citet{RandZ15} formalize another Hoare logic for
probabilistic programs using the \textsf{Coq} proof assistant. \citet{ellora}
implement a general-purpose logic in the \textsf{EasyCrypt} framework, and
verify a representative set of randomized algorithms. \citet{KKMO16}
develop a weakest precondition logic to reason about expected run-time
of probabilistic programs. 

Most of these works support general probabilistic reasoning and
additional features like non-determinism, so they most likely could
formalize the examples that we consider.
However, our logic \SYSTEM aims at a sweet spot in the design
space, combining expressivity with simplicity of the assertion language.
The design of \SYSTEM is inspired by existing
\emph{relational} program logics, such
as \Sprhl~\citep{BartheGZ09} and \Saprhl~\citep{BartheKOZ12}.  These
logics support rich proofs about probabilistic
properties with purely non-probabilistic assertions, using a
powerful coupling abstraction from probability
theory~\citep{BartheEGHSS15} rather than the union bound.

Finally, there are many algorithmic techniques for verifying
probabilistic programs. Probabilistic model-checking is a
successful line of research that has delivered mature and
practical tools and addressed a broad range of case studies;
\citet{KwiatkowskaNP02,Katoen08,BaierK11} 
cover some of the most interesting developments in the field.
Abstract interpretation of probabilistic programs is another
rich source of techniques; see e.g.\, \citet{Monniaux00,CousotM12}.  
\citet{KatoenMMM10} infer linear invariants for the \textsf{pGCL} language 
of~\citet{Morgan96}. There are several approaches based on martingales for
reasoning about probabilistic loops; \citet{ChakarovS13,ChakarovS14} use
martingales for inferring expectation
invariants, while
\citet{FioritiH15} use martingales for analyzing probabilistic
termination. \citet{SampsonPMMGC14} use a mix of static and dynamic
analyses to check probabilistic assertions for probabilistic programs.

\section{Conclusion and perspective}
We propose \SYSTEM, a lightweight probabilistic Hoare logic
based on the union bound. Our logic can prove properties about bad
events in cryptography and accuracy of differentially private
mechanisms. Of course, there are examples
that we cannot verify. For instance, reasoning involving independence
of random variables, a common tool when analyzing randomized
algorithms, is not supported.
Accordingly, a natural next step is to explore
logical methods for reasoning about independence, or to embed \SYSTEM
into a more general system like \textsf{pGCL}.

\subparagraph*{Acknowledgements}

This work was partially supported by NSF grants TWC-1513694,
CNS-1065060 and CNS-1237235, by EPSRC grant EP/M022358/1 and
by a grant from the Simons Foundation ($\#360368$ to Justin Hsu).

\bibliographystyle{abbrvnat}% the recommended bibstyle
\bibliography{header,refs}

\iffull
% --------------------------------------------------------------------
\appendix\clearpage

This appendix details the missing definitions of this paper main body,
along with the proof of soundness of the presented logic.

\section{A bit more on discrete distributions}

We start by defining some standard sub-distributions that are needed
for giving the denotation semantic of our language:

\begin{definition}
Let $T$ be some set and $x \in T$. We denote by
$\munit{x}^T \in \Dist(T)$ (resp. $\mnull^T \in \Dist(T)$) the Dirac
distribution over $T$ and centered on $x$ (resp. the null
sub-distribution over $T$):

\begin{center}
  $\begin{array}{l@{\,}l@{\hspace{1cm}}l@{\,}l}
    \munit{x}^T &= \lambda v \in T .\, \left\{
      \begin{array}{@{}l@{\hspace{.5cm}}l}
        1 & \mbox{if $x = v$} \\
        0 & \mbox{otherwise}
      \end{array} \right. &
    \mnull^T &= \lambda v \in T .\, 0 \\
   \end{array}$
\end{center}

We write $\munit{x}$ and $\mnull$, stripping $T$, when it is clear
from the context.

Let $T$ and $U$ be two sets. We denote by $\mlet{x}{\mu}{E(x)}$,
where $\mu$ is a sub-distribution over $T$ and
$E : T \rightarrow \Dist(U)$, the sub-distribution with mass function
$\lambda v .\, \sum_{x \in T} E(x)(v)\ \mu(x)$.
\end{definition}

It is convenient to introduce the notion of restriction of a distribution.

\begin{definition}[Restriction of a sub-distribution]
  Let $\mu$ be a sub-distribution over $T$, and let $P$ be a predicate
  over $T$. Then, the \emph{restriction} of $\mu$ to $P$ is defined as

  \begin{center}
  $\mu_{| P}(x) \triangleq
      \begin{cases}
        \mu(x) & \mbox{if $P(x)$} \\
        0 & \mbox{otherwise.}
      \end{cases}$
  \end{center}

  From the definition, it is clear that
  $\Pr_{\mu{| P}}[Q] = \Pr_{\mu}[P \land Q]$.
\end{definition}

\section{Denotational semantics}

We now give the denotation semantics of our language. We start by
interpreting the expressions and distribution expressions, and then
move to the interpretation of commands.

\subsection{Types, expressions and distribution expressions}

We fix a set $\Types = \{ \tau, \sigma, \ldots \}$ of types.  We
assume that $\Types$ contains at least the unit type ($\tunit$), along
with the types for booleans ($\tbool$) and integers ($\tint$). For a variable
$x \in \Vars$, we denote the type associated to $x$ by
$\varty{x}$. Moreover, for $\tau \in \Types$, we write $\Vars_\tau$
for the subset $\{ x \in \Vars \mid \varty{x} = \tau \}$ of $\Vars$,
and require that it is infinite.

We also assume given a set $\Ops$ of operators and $\DOps$ of
distribution operators.  To each operator $o \in \Ops$ is associated
an arity $o : [\tau_i]_{i \le n} \rightarrow \tau$, where $[\tau_i]_i$
is the domain of $o$ and $\tau$ its codomain. Likewise, to each
distribution operator $d \in \DOps$ is associated an arity
$d : \tau \rightarrow \sigma$, meaning that $d$ is a distribution over
$\sigma$ parameterized by a value of type $\tau$.

\medskip

We can now give the syntax of expressions and distribution
expressions:

\begin{definition}[Expressions \& distribution expressions]
  The set of expressions of type $\tau$, written $\Expr_\tau$, is
  defined by:

  \begin{center}
    $\Expr_\tau ::= x \in \Vars_\tau \mid o(e_1,\ldots,e_n)$
    $\quad\quad$ with
    $o : [\tau_i]_i \rightarrow \tau$ and $\forall i.\, e_i \in \Expr_{\tau_i}$.
  \end{center}

  Likewise, the set of distribution expressions over $\sigma$ is
  defined by:

  \begin{center}
    $\DExpr_\tau ::= d(e)$ $\quad\quad$ with
    $d : \sigma \rightarrow \tau$ and $e \in \Expr_\sigma$.
  \end{center}
\end{definition}

\medskip

We now move to the interpretation of types, expressions and
distribution expressions.

\subsection{Interpretation of types}

For any type $\tau$, the set $\denot{\tau}$ denotes the interpretation
of $\tau$: $\denot{\tunit} = \{ \bullet \}$,
$\denot{\tbool} = \{ \True, \False \}$, $\denot{\tint} = \mathbb{Z}$ and
$\denot{\treal} = \mathbb{R}$.

\subsection{Interpretation of expressions}

For any operator $o$ with arity $[\tau_i]_i \rightarrow \tau$, we
assume given
$\osem{o} : \bigtimes_i \denot{\tau_i} \rightarrow \denot{\tau}$.
The interpretation of an expression $e$ w.r.t a \emph{typed} valuation
$\rho : \Pi (x : \Vars) .\, \denot{\varty{x}}$ (i.e. w.r.t. a function
that associates a value in $\denot{\varty{x}}$ to any variable
$x \in \Vars$) is defined as usual: $\denot{x}_\rho = \rho(x)$ and
$\denot{o(e_1,\ldots,e_n)}_\rho =
\osem{o}(\denot{e_1}_\rho,\ldots,\denot{e_n}_\rho)$.
If $e \in \Expr_\tau$, we have that $\denot{e}_\rho \in \denot{\tau}$.

Likewise, for any distribution operator $d : \sigma \to \tau$ is
associated a function $\osem{d}$ from $\denot{\sigma}$ to
$\Dist(\denot{\tau})$. The interpretation of a distribution expression
$d(e)$ w.r.t.\ a valuation $\rho$, written $\denot{d(e)}_\rho$, is defined
by $\denot{d(e)}_\rho = \osem{d}(\denot{e}_\rho)$.

\subsection{Denotational semantics}

A memory $m$ is any map of type
$\Pi (x : \Vars \cup \{ \tadv \} ) .\, \denot{\varty{x}}$, where
$\tadv$ is a special variable dedicated to the storage of the (shared)
state of the external procedures --- associating an abstract type
$\varty{\tadv} = \tAdv$ to it. Note that memories can be considered as
valuations, simply forgetting the binding for $\tadv$.
For any external procedure $\Adv$ taking a parameter of type $\tau$
and returning a value of type $\sigma$, we assume given an
interpretation
$\osem{\Adv} : \tAdv \times \denot{\tau} \rightarrow \Dist(\tAdv \times
\denot{\sigma})$.

Finally, if $f$ is a internal procedure, we denote by $\farg{f}$
(resp.\ $\fbody{f}$, $\fret{f}$) the argument name (resp.\ the body,
the return expression) of $f$.

\begin{definition}[Denotational Semantics]
  The denotational semantics of a command maps a memory to a
  sub-distribution over memories and is given in
  Figure~\ref{fig:semantic}, where $\Fail$ is an extra command that
  never returns, and $(\Cond{e}{c})^n_\perp$ is inductively defined by:

  \begin{center}
  $\begin{array}{l@{\,}l}
    (\Condt{e}{c})^0_\perp &= \Condt{e}{\Fail} \\
    (\Condt{e}{c})^{n+1}_\perp &= \Condt{e}{\{ \Seq{c}{(\Condt{e}{c})^n_\perp} \}} \\
  \end{array}$
  \end{center}
\end{definition}

\begin{figure}[h!b]
\begin{center}
$\begin{array}{r@{\,}c@{\,}l}
\denot{\Fail} & = &
  \lambda m .\, \mnull \\
\denot{\Skip} & = &
  \lambda m .\, \munit{m} \\
\denot{\Ass{x}{e}} & = &
  \lambda m .\, \munit{m\udp{x}{\denot{e}_m}} \\
\denot{\Rand{x}{e}} & = &
  \lambda m .\, \mlet{v}{\denot{e}_m}{\munit{m\udp{x}{v}}} \\
\denot{\Seq{c1}{c2}} & = &
  \lambda m .\, \mlet{\xi}{\denot{c_1}(m)}{\denot{c_2}(\xi)} \\
\denot{\Cond{e}{c1}{c2}} & = &
  \lambda m .\, \left\{
    \begin{array}{@{}ll}
      \denot{c_1}(m) & \mbox{ if $\denot{e}_m = \True$} \\
      \denot{c_2}(m) & \mbox{ if $\denot{e}_m = \False$} \\
    \end{array}\right. \\
\denot{\WWhile{e}{c}} & = &
  \lambda m .\, \lim_{n\infty}
    \denot{(\Condt{e}{c})^n_\perp}(m) \\
\denot{\Call{x}{\Adv}{e}} & = &
  \lambda m .\,
    \mlet{(v_\tadv, v)}{\osem{\Adv}(m(\tadv), \denot{e}_m)}%
         {\munit{m\udp{\tadv}{v_\tadv}\udp{x}{v}}} \\
\denot{\Call{x}{f}{e}} & = &
  \lambda m .\, \denot{\Ass{\farg{f}}{e}; \fbody{f}; \Ass{x}{\fret{f}}}(m) \\
  & = &
  \lambda m .\, \mlet{\xi}{\denot{\fbody{f}}(m')}
    {\munit{\xi\udp{x}{\xi(\fret{f})}}} \\
  && \mbox{where $m' = m\udp{\farg{f}}{\denot{e}_m}$} \\
\end{array}$
\end{center}

\caption{\label{fig:semantic} Denotational Semantics}
\end{figure}

The following lemma is useful in the proof of soundness.

\begin{lemma}\label{l:cmodfv}
  Let $c$ be a command, $m_1$, $m_2$ be two memories s.t.
  $m_2 \in \supp(\denot{c}(m_1))$. Then, $\forall x \in \Var$ that is
  not written by $c$ and $m_1[x] = m_2[x]$.
\end{lemma}

\begin{proof}
  By a direct induction on the structure of $c$.
\end{proof}

Before moving to the soundness proof, we make clear what it means for
a command to modify a variable

\begin{definition}
  Let $c$ be command. The set $\cmod(c) \subseteq \Vars$ of program
  variables modifies by $c$ is defined by induction on the structure
  of $c$:

  \begin{center}
    $\begin{array}{@{\cmod(}l@{)\,}l}
      \Fail &= \emptyset \\
      \Skip &= \emptyset \\
      \Ass{x}{e} &= \{ x \} \\
      \Rand{x}{d(e)} &= \{ x \} \\
      \Seq{c_1}{c_2} &= \cmod(c_1) \cup \cmod(c_2) \\
      \Cond{e}{c_1}{c_2} &= \cmod(c_1) \cup \cmod(c_2) \\
      \WWhile{e}{c} &= \cmod(c) \\
      \Call{x}{\Adv}{e} &= \{ x \} \\
      \Call{x}{f}{e} &= \{ x, \farg{f} \} \cup \cmod(\fbody{f}) \\
    \end{array}$
  \end{center}
\end{definition}

Note that the expressions do not have side effects.

\section{Soundness proof}

The logical entailment relation $m \models \Phi$ being abstract, we
assume here the following property:

\begin{lemma}
  Let $m$, $\Phi$ and $e$ s.t. $m \models \Phi[e/x]$. Then,
  $m\udp{x}{e} \models \Phi$.
\end{lemma}

Note that, beside the former property, $m \models \Phi$ gives a
standard model to first order connectives.
We can now prove Theorem~\ref{th:soundness}:

\begin{proof}[Proof of Theorem~\ref{th:soundness}]
  The proof is by induction on the derivation of
  $\ahl{c}{\Phi}{\Psi}{\beta}$, by case analysis on the last rule:

  \begin{description}
  \setlength{\itemsep}{.5em}

  \item[\rname{Skip}] We have $\Psi \equiv \Phi$ and $\beta = 0$. Let
    $m \models \Phi$. Then
    $\Pr_{\denot{\Skip}(m)}[\neg \Psi] = \Pr_{\munit{m}}[\neg \Psi] =
    \sum_\xi \left({\munit{m}}\right)_{|\neg \Psi}(\xi) =
    \left({\munit{m}}\right)_{|\neg \Psi}(m) = 0$,
    the last equality being a direct consequence of
    $m \models \Phi (\equiv \Psi)$.

  \item[\rname{Weak}] We have $\ahl{c}{\Phi'}{\Psi'}{\beta'}$ with
    \begin{enumerate*}[i)]
      \item $\models \Phi \implies \Phi'$,
      \item $\models \Psi' \implies \Psi$, and
      \item $\beta' \le \beta$.
    \end{enumerate*}
    Let $m \models \Phi$. By i), $m \models \Phi'$. Hence, by
    induction hypothesis, $\Pr_{\denot{c}(m)}[\neg \Psi'] \le \beta'$.
    From ii) \& iii),
    $\Pr_{\denot{c}(m)}[\neg \Psi] \le \Pr_{\denot{c}(m)}[\neg \Psi']
    \le \beta' \le \beta$.

  \item[\rname{Seq}] We have $\ahl{c_1}{\Phi}{\Xi}{\beta_1}$ and
    $\ahl{c_2}{\Xi}{\Psi}{\beta_2}$, with $c \equiv \Seq{c_1}{c_2}$
    and $\beta = \beta_1 + \beta_2$. Let $m \models \Phi$. Then,

    \begin{equation*}
    \begin{split}
    \Pr_{\denot{c}(m)}[\neg \Psi]
      & = \sum_{\xi \models \neg\Psi} \denot{\Seq{c_1}{c_2}}(m)(\xi)
        = \sum_{\xi \models \neg\Psi}\left[
            \mlet{\theta}{\denot{c_1}(m)}{\denot{c_2}(\theta)}
          \right](\xi) \\[.35em]
      & \quad\quad\quad \mbox{(swap $\sum$ --- inner $\sum$ being hidden in $\int$)} \\
      & = \sum_\theta \sum_{\xi \models \neg\Psi}
            \denot{c_2}(\theta)(\xi) \cdot \denot{c_1}(m)(\theta) \\[.35em]
      & \quad\quad\quad \mbox{(split external $\sum$ on $\models \Xi$ and refold inner $\Pr$)} \\
      & = \sum_{\theta \models \Xi}
            \denot{c_1}(m)(\theta) \cdot
              \underbrace{\Pr_{\denot{c_2}(\theta)}[\neg\Psi]}_{\le \beta_2} +
          \sum_{\theta \models \neg\Xi}
            \denot{c_1}(m)(\theta) \cdot
              \underbrace{\Pr_{\denot{c_2}(\theta)}[\neg\Psi]}_{\le 1} \\
      & \le
          \underbrace{\left(
              \sum_{\theta \models \Xi} \denot{c_1}(m)(\theta)
          \right)}_{= \Pr_{\denot{c_1}(m)}[\Xi] \le 1} \cdot \beta_2 +
          \underbrace{%
            \sum_{\theta \models \neg\Xi} \denot{c_1}(m)(\theta)%
          }_{= \Pr_{\denot{c_1}(m)}[\neg\Xi] \le \beta_1}\\
      & \le \beta_1 + \beta_2\\
    \end{split}
    \end{equation*}

  \item[\rname{Assn}] We have $c \equiv \Ass{x}{e}$,
    $\phi \equiv \Xi[e/x]$, $\psi \equiv \Xi$ and $\beta = 0$. Let
    $m \models \Xi[e/x]$. By substitutivity of the logical entailment
    relation, $m\udp{x}{\denot{e}_m} \models \Xi$. Then,
    $\Pr_{\denot{c}(m)}[\neg \Xi] = \sum_\xi \left[(
      \munit{m\udp{x}{\denot{e}_m}}
    )_{| \neg \Xi}\right](\xi)
    = \left[(\munit{m\udp{x}{\denot{e}_m}})_{| \neg \Xi}\right](m\udp{x}{\denot{e}_m})
    = 0$.

  \item[\rname{Rand}] The premise directly implies the conclusion ---
    the rule is semantical.

  \item[\rname{If}] We have
    $\ahl{c_{\True}}{\Phi \land e}{\Psi}{\beta}$ and
    $\ahl{c_{\False}}{\Phi \land \neg e}{\Psi}{\beta}$, with
    $c \equiv \Cond{e}{c_{\True}}{c_{\False}}$. Let $m \models \Phi$
    and let $b = \denot{e}_m$.  Then, $m \models \Phi \land e = b$,
    and by application of the induction hypothesis,
    $\Pr_{\denot{c}(m)}[\neg \Psi] = \Pr_{\denot{c_b}(m)}[\neg \Psi]
    \le \beta$.

  \item[\rname{While}] We have $c \equiv \WWhile{e}{c_I}$,
    $\Phi \equiv I \land e_v \le k$, $\Psi \equiv I \land \neg e$ and
    $\beta = k \cdot \beta_I$, with $\ahl{c_I}{I}{I}{\beta_I}$,
    $\models I \to (e_v \leq k) \land (e_v \leq 0 \to \neg e)$ and
    $\forall \eta > 0 .\, \ahl{c_I}{I \land e \land e_v = \eta}{e_v <
      \eta}{0}$.
    The proof is done by (strong) induction on $k$. Let
    $m \models I \land e_v \le k$.

    If $m \models \neg e$, then $(\Cond{e}{c_I})^t_\perp\ (m) = \munit{m}$
    for any $t$. Hence, $\denot{\WWhile{e}{c_I}}(m) = \lim_{n\infty} \munit{m} =
    \munit{m}$; and, from $m \models I \land \neg e$, we have
    $\Pr_{\denot{c}(m)}[\neg (I \land \neg e)] =
    \Pr_{\munit{m}}[\neg (I \land \neg e)] =
    (\munit{m})_{| \neg (I \land \neg e)}\ (m) = 0 \le k \cdot \beta_I$.

    Otherwise, $m \models e$. From the logical premises, we have
    $m \models 0 < e_v \le k$. Moreover,
    $\denot{(\Cond{e}{c_I})^{t+1}_\perp}(m)
     = \denot{\Seq{c_I}{(\Cond{e}{c_I})^t_\perp}}(m)$,
    and thus, $\denot{c}(m) = \denot{\Seq{c_I}{c}}(m)$. By a reasoning
    similar to the one of \rname{Seq}, using $I \land e_v \le (k-1)$ as
    the intermediate assertion, it suffices to show that
    \begin{enumerate*}[i)]
    \item \label{snd:while:b1} $\Pr_{\denot{c_I}(m)}[\neg (I \land e_v \le (k-1))]
      \le \beta_I$, and
    \item \label{snd:while:b2} for any $\xi \models I \land e_v \le (k-1)$,
      $\Pr_{\denot{c}(\xi)}[\neg (I \land \neg e)] \le (k-1) \cdot \beta_I$.
    \end{enumerate*}
    Point~\ref{snd:while:b2}) is obtained by an application of the inner
    induction hypothesis. For Point~\ref{snd:while:b1}), we have:

    \begin{tightcenter}
      $\Pr_{\denot{c_I}(m)}[\neg (I \land e_v \le (k-1))]
        \le \underbrace{\Pr_{\denot{c_I}(m)}[\neg I]}_{\le\ \beta_I}
        +   \underbrace{\Pr_{\denot{c_I}(m)}[\neg (e_v \le (k-1))]}_{=\ 0}$
    \end{tightcenter}

    \noindent the comparison to $\beta_I$ coming form
    $\ahl{c}{I}{I}{\beta_I}$, whereas the comparison to $0$ is a
    direct consequence of
    $m \models I \land e \land e_v = k \to e_v < k$, obtained by
    instantiation of the logical premises.

  \item[\rname{Ext}] We have $c \equiv \Call{x}{\Adv}{e}$,
    $\Phi \equiv \forall v .\, \Psi[v/x]$ and $\beta = 0$. Let
    $m \models \Phi$. Then,

    \begin{equation*}
    \begin{split}
     \Pr_{\denot{c}(m)}[\neg \Psi]
       & = \sum_{\xi \models \neg \Psi} \sum_{(v_\tadv, v)} \left[
             \munit{\underbrace{m\udp{\tadv}{v_\tadv}\udp{x}{v}}_{\alpha(v_\tadv, v)}}
             \cdot \osem{\Adv}(m[\tadv], \denot{e}_m)(v_\tadv, v)
           \right](\xi) \\
       & = \sum_{(v_\tadv, v)} \sum_\xi \left[
             \munit{\alpha(v_\tadv, v)} \cdot
             \underbrace{\osem{\Adv}(m[\tadv], \denot{e}_m)(v_\tadv, v)}_{\le 1}
           \right]_{| \neg \Psi} (\xi) \\
       & \le \sum_{(v_\tadv, v)} \sum_\xi \left[
               \munit{\alpha(v_\tadv, v)}
             \right]_{| \neg \Psi} (\xi)
         = \sum_{(v_\tadv, v)}  \underbrace{\left[
               \munit{\alpha(v_\tadv, v)}
             \right]_{| \neg \Psi} (\alpha(v_\tadv, v))}_{=\ 0} \\
       & = 0
    \end{split}
    \end{equation*}

    \noindent noticing that $m \models \forall v .\, \Psi[v/x]$ with
    $\tadv \notin \Psi$ implies $\alpha(v_\tadv, v) \models \Psi$.

  \item[\rname{Call}] This is a direct consequence of the properties
    for \rname{Seq} \& \rname{Assn}.

  \item[\rname{Frame}] We have $\Phi \equiv \Psi$, $\beta = 0$ and $c$
    does not modify the variables free in $\Phi$. Then,

    \begin{equation*}
    \begin{split}
     \Pr_{\denot{c}(m)}[\neg \Psi]
      & = \Pr_{\denot{c}(m)}[\neg \Phi]
        = \sum_{\xi \models \neg \Phi} \denot{c}(m)(\xi)
        = \sum_{\xi \in \supp(\denot{c}(m))} (\denot{c}(m))_{| \neg \Phi} (\xi).
    \end{split}
    \end{equation*}

    From Lemma~\ref{l:cmodfv}, for any $\xi \in \supp(\denot{c}(m))$,
    considering that $c$ does not modify the free variables
    $\FV(\Phi)$ of $\Phi$, we have
    $m_{| \FV(\phi)} = \xi_{| \FV(\phi)}$. Hence, $m \models \Phi$
    implies $\xi \models \Phi$, and:

    \begin{center}
      $\displaystyle\sum_{\xi \in \supp(\denot{c}(m))}
         \underbrace{(\denot{c}(m))_{| \neg \Phi} (\xi)}_{=\ 0} = 0$.
    \end{center}

  \item[\rname{And}] We have $\ahl{c}{\Phi}{\Psi_1}{\beta_1}$ and
    $\ahl{c}{\Phi}{\Psi_2}{\beta_2}$ with
    $\Psi \equiv \Psi_1 \land \Psi_2$ and
    $\beta = \beta_1 + \beta_2$.
    Let $m \models \Phi$. Then, by induction hypothesis,
    $\Pr_{\denot{c}(m)}[\neg \Phi_1] \le \beta_1$ and
    $\Pr_{\denot{c}(m)}[\neg \Phi_2] \le \beta_2$. Hence,
    $\Pr_{\denot{c}(m)}[\neg (\Phi_1 \land \Phi_2)] =
     \Pr_{\denot{c}(m)}[\neg \Phi_1] + \Pr_{\denot{c}(m)}[\neg \Phi_2] -
     \Pr_{\denot{c}(m)}[\neg \Phi_1 \land \neg \Phi_2] \le
    \Pr_{\denot{c}(m)}[\neg \Phi_1] + \Pr_{\denot{c}(m)}[\neg \Phi_2]
    \le \beta_1 + \beta_2 = \beta$.

  \item[\rname{Or}] We have $\ahl{c}{\Phi_1}{\Psi}{\beta}$ and
    $\ahl{c}{\Phi_2}{\Psi}{\beta}$ with
    $\Phi \equiv \Phi_1 \lor \Phi_2$. Let $m \models \Phi$. Then,
    $m \models \Phi_1$ or $m \models \Phi_2$. W.l.o.g. we can assume
    $m \models \Phi_1$. We then obtain the expected result
    $\Pr_{\denot{c}(m)}[\neg \Psi] \le \beta$ by application of the
    induction hypothesis.

  \item[\rname{False}] We have $\Psi \equiv \False$ and
    $\beta = 1$. Let $m \models \Phi$.  Then
    $\Pr_{\denot{c}(m)}[\neg \Psi] \le 1 = \beta$. \qedhere
  \end{description}

\end{proof}

%%% Local Variables:
%%% mode: latex
%%% TeX-master: "../main"
%%% End:

\fi

\end{document}